\documentclass[aps,pra,reprint,superscriptaddress,nofootinbib]{revtex4-2}

\usepackage{graphicx}
\usepackage{subcaption}
\usepackage{hyperref}
\usepackage{xcolor}
\usepackage{amsmath}
\usepackage{amssymb}
\usepackage{tikz}
\usetikzlibrary{decorations.pathmorphing, arrows.meta}
\usepackage{bm}
\usepackage{comment}
\usepackage{etoolbox}
\usepackage{ragged2e}
\usepackage{physics}
\usepackage{float}

\makeatletter
\long\def\@makecaption#1#2{%
  \vskip\abovecaptionskip
  \justifying \small
  \sbox\@tempboxa{#1: #2}%
  \ifdim \wd\@tempboxa >\hsize
    #1: #2\par
  \else
    \global \@minipagefalse
    \box\@tempboxa
  \fi
  \vskip\belowcaptionskip
}
\makeatother

\newcommand{\mbf}{\boldsymbol}

\begin{document}

\title{Subradiant collective states for precision sensing via transmission spectra}

\author{Diego Zafra-Bono}
\affiliation{Department
of Physics, Universitat Politècnica de Catalunya, Barcelona 08034,
Spain}
\affiliation{Department of Physics, Harvard University, Cambridge, Massachusetts 02138, USA}

\author{Oriol Rubies-Bigorda}
\affiliation{Physics Department, Massachusetts Institute of Technology, Cambridge, Massachusetts 02139, USA}
\affiliation{Department of Physics, Harvard University, Cambridge, Massachusetts 02138, USA}

\author{Susanne F. Yelin}
\affiliation{Department of Physics, Harvard University, Cambridge, Massachusetts 02138, USA}

\date{\today}

\begin{abstract}
When an ensemble of quantum emitters interacts with a common radiation field, their emission becomes collective, giving rise to superradiant and subradiant states, characterized by broadened and narrowed linewidths. In this work, we propose to harness subradiant states for quantum metrology; such states naturally arise in subwavelength-spaced atomic arrays in free space and in small ensembles of emitters coupled to one-dimensional waveguides. We demonstrate that their collective optical response yields sharp, narrow features in the transmittance spectrum, which can be used to enhance sensitivity to external perturbations. This improved sensitivity can be applied to atomic clock operation, spatially resolved imaging of emitter positions, and enables precise detection of both global and spatially varying detunings—such as those induced by electromagnetic fields or gravitational gradients.
\end{abstract}

\maketitle

\section{\label{sec:intro}Introduction}

Ensembles of quantum emitters have emerged in recent years as promising platforms for controlling light–matter interactions at the quantum level. In particular, when emitters couple to a common radiation field, they no longer behave as independent scatterers but instead scatter light collectively~\cite{dicke_coherence_1954, cohen-tannoudji_atom-photon_1998, meystre_elements_2007, lambropoulos_fundamentals_2007, steck_quantum_2007}. The light-mediated interactions between emitters lead to the emergence of superradiant and subradiant states, which respectively exhibit enhanced and suppressed photon emission rates due to the constructive or destructive interference of decay paths~\cite{dicke_coherence_1954, rehler_superradiance_1971, gross_superradiance_1982, asenjo-garcia_exponential_2017}. The precise nature of these collective effects is strongly influenced by the electromagnetic environment. For emitters coupled to a waveguide, the guided modes mediate long-range interactions~\cite{asenjo_1D_2017}, and even two emitters are sufficient to support nearly perfect subradiant (dark) states that decouple from the propagating field~\cite{gonzalez-tudela_entanglement_2011, lalumiere_input-output_2013}. In free space, dipole–dipole interactions between emitters decay rapidly with distance and become negligible after only a few wavelengths. However, in periodic arrays with lattice spacing smaller than the optical transition wavelength, cooperative phenomena remain strong~\cite{asenjo-garcia_exponential_2017} and lead to highly structured optical responses, such as directional emission and light scattering~\cite{bettles_enhanced_2016, shahmoon_cooperative_2017, moreno-cardoner_subradiance-enhanced_2019,rui_subradiant_2020}. These arrays also host dark collective excitations delocalized across the entire system, which can be harnessed for photon storage and routing~\cite{asenjo-garcia_exponential_2017, facchinetti_storing_2016, manzoni_optimization_2018,rubies-bigorda_photon_2022,Ruostekoski_memory} or for mediating long-range interactions~\cite{masson_atomic-waveguide_2020, patti_controlling_2021,Shah2024}.

Subradiant states, characterized by their long lifetimes and narrow linewidths, are natural candidates for quantum metrology. Quantum metrology aims to exploit quantum resources to enhance measurement precision~\cite{giovannetti_quantum_2006, giovannetti_advances_2011}, with atomic clocks serving as a paradigmatic example, where ultra-narrow optical transitions enable high-precision frequency standards~\cite{ludlow_optical_2015}. Atomic arrays have recently been explored as a metrological platform, for instance by studying the influence of collective energy shifts and decay rates on Ramsey interrogation sequences~\cite{array_clock_1,array_clock_2}, or by probing population exchange between impurities embedded in subwavelength lattices~\cite{OliverStefan_sensing}.

\begin{figure}[H]    \includegraphics[width=0.89\linewidth]{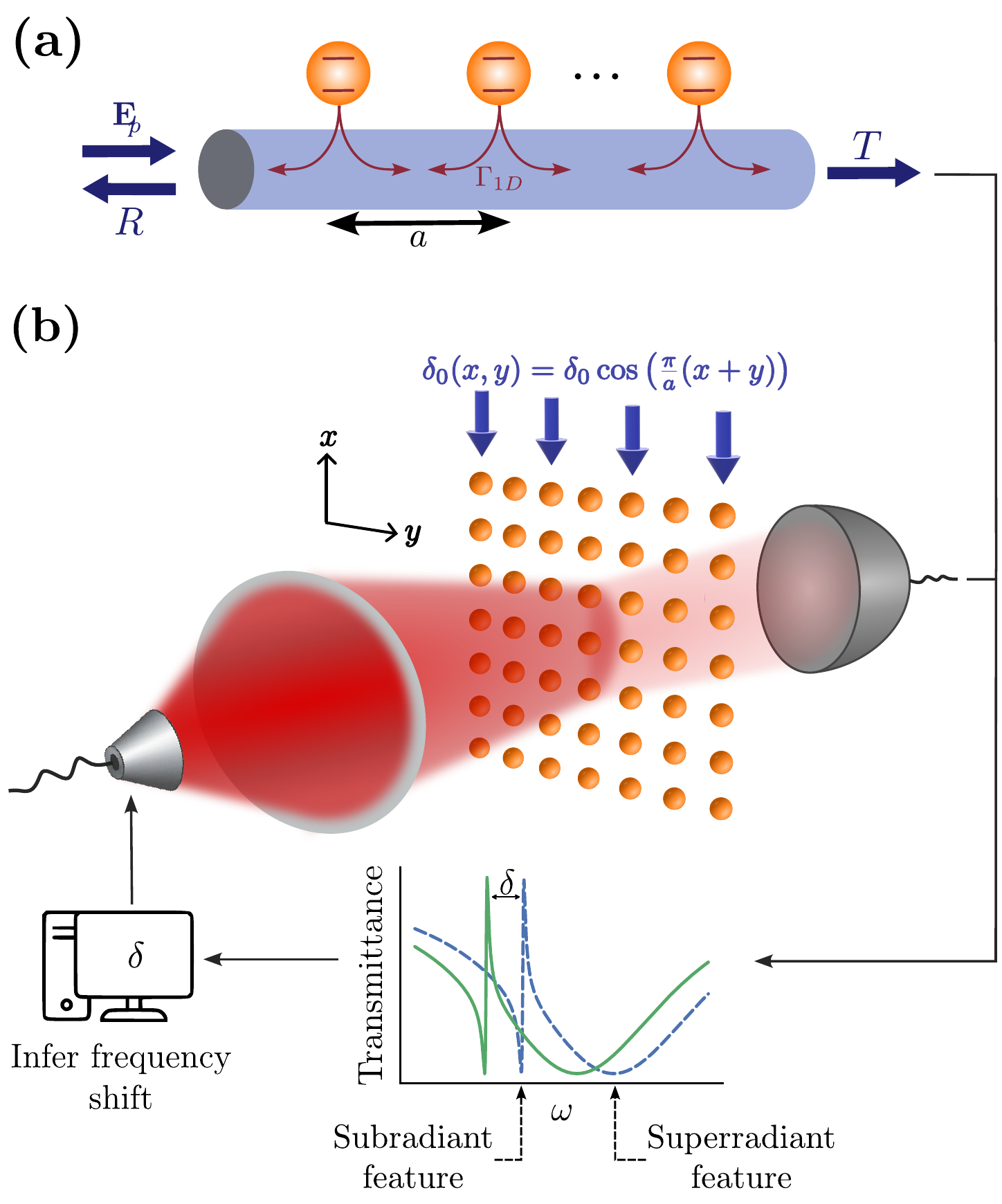}
    \centering
    \caption{\textbf{Experimental setups}. Schematic of metrological protocols based on collective light scattering. A weak probe laser illuminates (a) a one-dimensional array of emitters coupled to a waveguide, or (b) a two-dimensional atomic array in free space with a spatially-varying atomic frequency shift $\delta_0(x,y)$. The measured transmittance depends on the collective scattering properties of the emitters. Enhanced sensitivity arises from the narrowed linewidth of subradiant states. The system allows detection of frequency shifts induced by environmental perturbations and can operate as an atomic clock, where changes in the laser frequency manifest as a shift in the transmittance spectrum.
    In the latter case, the detuning $\delta$ inferred from the transmittance data is used in a feedback loop to adjust and stabilize the laser frequency.}
    \label{fig:setup}
\end{figure}

Although subradiant states offer significant potential for precision metrology, their practical deployment has been limited: the very radiative protection that gives rise to their long lifetimes also hinders their accessibility, and their resonances depend intricately on the specific collective mode, making them challenging to characterize. Subradiant modes have nonetheless been harnessed in platforms such as plasmonic nanostructures, where their narrow resonances enable biosensing applications~\cite{yanik_seeing_2011}.
In the context of neutral atoms, Facchinetti and Ruostekoski theoretically showed that Zeeman shifts allow access to ultra-narrow subradiant modes in perfectly ordered planar lattices of multilevel atoms. These subradiant modes differ from the superradiant ones in their polarization profile, and give rise to phenomena akin to electromagnetically induced transparency that enable magnetometry with enhanced sensitivities~\cite{facchinetti_interaction_2018}.

In this work, we access the subradiant states of arrays of two-level atoms by introducing spatially varying frequency shifts (i.e. AC Stark shifts). This mechanism effectively couples subradiant modes to radiative ones and can be implemented in both free-space and waveguide-coupled arrays of emitters. The resulting subradiant states give rise to narrow spectral features~\cite{de_paz_bound_2023} that appear in transmission spectra [see Fig.~\ref{fig:setup}], enabling the detection of global frequency shifts and spatially varying perturbations, such as position-dependent detunings or atomic displacements. We further show that narrow spectral features valuable for metrology emerge in waveguide-coupled arrays even in the absence of control frequency shifts. Finally, we assess the robustness of the proposed scheme against typical experimental imperfections and estimate the achievable frequency precision.

The remainder of this paper is organized as follows. In Sec.~\ref{sec:model}, we present the theoretical framework of our protocol. Section~\ref{sec:global_detuning} discusses how subradiant states enable the detection of global detunings, either through direct population of subradiant states in waveguide-coupled arrays near the Dicke limit, or through indirect excitation via coupling to superradiant states when subradiant states are perfectly dark. In Sec.~\ref{sec:site-resolved_metrology}, we demonstrate how waveguide-coupled arrays enable site-resolved sensing of local detunings and atomic displacements. Finally, Sec.~\ref{sec:errors} examines the robustness of our protocol against experimental imperfections---including atomic motion, missing atoms, and imperfect waveguide coupling--- and provides an estimate of its performance, demonstrating that subradiant states can still provide enhanced sensitivity under realistic conditions.

\section{System and model}\label{sec:model}

We consider an array of two-level atoms with transition frequency $\omega_0 = 2\pi c / \lambda_0$ and lowering operators $\hat{\sigma}_i = |g_i\rangle \langle e_i|$. The atoms are driven by a classical monochromatic field $\hat{\mathbf{E}}_p$. In a frame rotating at the laser frequency $\omega_L$, the driving Hamiltonian reads
\begin{equation}
    \hat{H}_{\text{drive}} = -\hbar \Delta_L \sum_{i = 1}^N \hat{\sigma}_i^\dagger \hat{\sigma}_i - \hbar\sum_{i=1}^N \left[\Omega_i \hat{\sigma}_i^\dagger + \Omega_i^* \hat{\sigma}_i\right],
\end{equation}
where $\Delta_L = \omega_L - \omega_0$ is the detuning of the laser from the atomic transition, and $\Omega_i = \left(\mathbf{d}^* \cdot \mathbf{E}^+_p(\mathbf{r}_i)\right)/\hbar$ is the Rabi frequency of atom $i$, located at $\mathbf{r}_i$ and with dipole matrix element $\mathbf{d}$. We also account for potential atom-dependent frequency shifts $\delta_i$ of the atomic transition frequency, with $|\delta_i| \ll \omega_0$, described by the Hamiltonian
\begin{equation}
            \hat{H}_{\text{detuning}} = -\hbar \sum_{i = 1}^N \delta_i \hat{\sigma_i}^\dagger \hat{\sigma}_i.
\end{equation}

The atoms are coupled to a shared vacuum electromagnetic environment via their dipole moments. Within the Born-Markov approximation, the photonic degrees of freedom can be traced out, yielding a master equation for the reduced density matrix $\hat{\rho}$ of the atomic system $\dot{\hat{\rho}} = -(i/\hbar)[\hat{{H}}, \hat{\rho}]+\mathcal{L}[\hat{\rho}]$. The Hamiltonian, $\hat{H} = \hat{H}_{\text{drive}}+\hat{H}_{\text{detuning}} + \hat{H}_{\text{dd}}$, contains three contributions: the driving term $\hat{H}_{\text{drive}}$, the detunings $\hat{H}_{\text{detunings}}$, and the coherent dipole-dipole interactions,
\begin{equation}
    \hat{H}_{\text{dd}} = \hbar \sum_{i,j=1}^N J_{ij} \hat{\sigma}_i^\dagger \hat{\sigma}_j,
\end{equation}
which describe 
coherent exchange of excitations between the atoms mediated by the vacuum electromagnetic field. The Lindblad superoperator $\mathcal{L}[\hat{\rho}]$ is given by~\cite{meystre_elements_2007,lambropoulos_fundamentals_2007,steck_quantum_2007,cohen-tannoudji_atom-photon_1998}
\begin{equation}\label{eq:Lindbladian}
    \mathcal{L}[\hat{\rho}] = \sum_{i,j=1}^N \frac{\Gamma_{ij}}{2} \left( 2 \hat{\sigma}_j \hat{\rho} \hat{\sigma}_i^\dagger - \hat{\sigma}_i^\dagger \hat{\sigma}_j \hat{\rho} - \hat{\rho} \hat{\sigma}_i^\dagger \hat{\sigma}_j \right).
\end{equation}
It describes the correlated decay of atomic excitations via photon emission, and thus accounts for collective spontaneous emission into the common electromagnetic environment.

The coherent $J_{ij}$ and dissipative $\Gamma_{ij}$ coupling coefficients between atoms $i$ and $j$ are determined by the electromagnetic environment and are given by~\cite{asenjo-garcia_exponential_2017}
\begin{equation}\label{eq:rates}
    J_{ij} - i \frac{\Gamma_{ij}}{2} = -\frac{\mu_0 \omega_0^2}{\hbar} \, \mathbf{d}^* \cdot \mathbf{G}(\mathbf{r}_i - \mathbf{r}_j, \omega_0) \cdot \mathbf{d},
\end{equation}
where $\mathbf{G}(\mathbf{r}_i - \mathbf{r}_j, \omega_0)$ is the dyadic Green's tensor describing the electric field at position $\mathbf{r}_i$ due to a point dipole located at $\mathbf{r}_j$. These couplings govern the collective emission and scattering of light, which occur through the eigenstates (or eigenmodes) $|\phi_\alpha\rangle$ of the non-Hermitian dipole-dipole Hamiltonian
\begin{equation}
\label{eq: effective_Ham}
    \hat{H}_\text{eff} = \sum_{i,j} \left( J_{ij} - i \frac{\Gamma_{ij}}{2} \right) \hat{\sigma}_i^\dagger \hat{\sigma}_j,
\end{equation}
with complex eigenvalues $\lambda_\alpha = J_\alpha - i \Gamma_\alpha / 2$. The real $J_\alpha$ and imaginary $\Gamma_\alpha$ parts respectively describe the energy shift and decay rate of each eigenmode. Modes with enhanced decay rates $\Gamma_\alpha > \Gamma_{jj}$ are referred to as \emph{superradiant}, whereas those with suppressed decay rates $\Gamma_\alpha < \Gamma_{jj}$ are termed \emph{subradiant}.

In this work, we explore how atomic arrays can be used to sense both global and local perturbations---such as uniform or position-dependent detunings $\delta_j$, as well as atomic displacements--- by measuring the steady-state transmittance of a laser drive incident on the atomic array. The positive-frequency component  of the total electric field at any position $\mathbf{r}$ is given by the superposition of the incident field $\hat{\mathbf{E}}_p^+(\mathbf{r})$ and the field scattered by the atoms,
\begin{equation}\label{eq:electric_field}
    \hat{\mathbf{E}}^+(\mathbf{r}) = \hat{\mathbf{E}}_p^+(\mathbf{r}) + \mu_0 \omega_0^2 \sum_{j = 1}^N  \mathbf{G}(\mathbf{r} - \mathbf{r}_j, \omega_0) \cdot \mathbf{d} \, \hat{\sigma}_j.
\end{equation}

For a laser field incident from the left of the array, we define the transmission coefficient as
\begin{equation}
    t = \frac{\langle \mathbf{E}^+(\mathbf{r}_\text{right}) \rangle}{\langle \mathbf{E}^+_p(\mathbf{r}_\text{right}) \rangle},
\end{equation}
where $\mathbf{r}_\text{right}$ denotes a position sufficiently far on the right side of the array, to avoid near-field effects. Similarly, the reflection coefficient $r$ is defined as the ratio between the amplitude of the reflected field (i.e., the field scattered to the left of the array) and the amplitude of the incident field. The transmittance and reflectance are given by $T = |t|^2$ and $R = |r|^2$.

The transmittance depends on the expectation value of the atomic coherences, $\langle \hat{\sigma}_j\rangle$, which can be readily obtained from the master equation as $d \langle \hat{\sigma}_j \rangle / dt = \text{Tr}\{ \dot{\hat{\rho}} \, \hat{\sigma}_j \}$. We focus on the weak-driving regime, such that the atomic system remains in the low-excitation limit with $\langle \sigma_i^\dagger \sigma_i \rangle \ll 1$ or $\langle \sigma_i^z\rangle \approx -1$ for all atoms $i$. In this limit, the steady-state atomic coherences satisfy the following linear system of equations
\begin{equation}\label{eq:steady_state}
    \left( - \Delta_L - \delta_j - i \frac{\Gamma_{jj}}{2} \right) \langle \hat{\sigma}_j \rangle + \sum_{i \neq j} \left( J_{ij} - i \frac{\Gamma_{ij}}{2} \right) \langle \hat{\sigma}_i \rangle = \Omega_j
\end{equation}
for all $j = 1,\dots,N$.

Throughout this work, we consider two distinct setups, illustrated in Fig.~\ref{fig:setup}:

\begin{enumerate}
\item \textbf{Atoms coupled to a 1D waveguide}. We first consider a linear chain of atoms coupled to a single-mode, one-dimensional waveguide. In this setting, the coherent and dissipative interactions between atoms are obtained from the Green's tensor for the waveguide $\mathbf{G}_{1D}(x, k_p)$ as~\cite{asenjo_1D_2017,fayard_many-body_2021}
    \begin{equation}
        (\mu_0 \omega_p^2 / \hbar) \, \mathbf{d}^* \cdot \mathbf{G}_{1D}(x, k_p) \cdot \mathbf{d} = \frac{i}{2} \Gamma_{1D} e^{i k_p |x|},
    \end{equation}
    where $\Gamma_{1D} \equiv \Gamma_{jj}$ is the decay rate of a single atom in the guided mode, and $\omega_p$ and $k_p$ are the frequency and wavenumber of the mode, respectively. It is worth noting that the light-mediated interactions in a waveguide extend over distances much larger than $\lambda$, enabling the formation of subradiant states even with only a few atoms.

\item \textbf{2D array in free space} The second setup consists of a square lattice of atoms in free space. The corresponding Green's tensor is given by
\begin{equation}
\begin{split}
    \mathbf{G}(\mathbf{r}, \omega_0) = \frac{e^{i k_0 r}}{4 \pi k_0^2 r^3} \Big[ (k_0^2 r^2 + i k_0 r - 1)\mathbb{I} \\
    + (-k_0^2 r^2 - 3i k_0 r + 3)\frac{\mathbf{r} \otimes \mathbf{r}}{r^2} \Big],
\end{split}
\end{equation}
where $r = |\mathbf{r}|$ and $k_0 = \omega_0 / c$ is the wavenumber of the atomic transition. The spontaneous decay rate of an atom in free space is given by $\Gamma_{jj} = \mu_0\omega_0^3|\mathbf{d}|^2/(3\pi\hbar c) \equiv \Gamma_0$. As opposed to the waveguide, the strength of the light-mediated interactions rapidly decays with distance.

In an infinite two-dimensional array, the translational symmetry of the lattice ensures that the non-Hermitian effective Hamiltonian in Eq.~(\ref{eq: effective_Ham}) is diagonal in the momentum basis within the single-excitation subspace (i.e. when a single excitation is shared among all atoms). The corresponding eigenstates are
\begin{equation}\label{eq:momentum_basis}
    \ket{\psi_{\mathbf{k}}} = \frac{1}{\sqrt{N}} \sum_{j=1}^N e^{i \mathbf{k} \cdot \mathbf{r}_j} \ket{e_j}.
\end{equation}
If the quasi-momentum $\mathbf{k}$ of the mode lies inside the light cone, i.e., $|\mathbf{k}| \leq \omega_0/c$, the mode can couple to propagating free-space photons and radiate; it is therefore termed superradiant (or bright). In contrast, states outside the light cone are perfectly subradiant (or dark), as they cannot phase-match to any propagating light mode.

\end{enumerate}

\section{Detection of global frequency shifts\label{sec:global_detuning}}

\begin{figure*}
\centering
\begin{subfigure}[b]{0.29\textwidth}
\centering
    \includegraphics[scale=0.73]{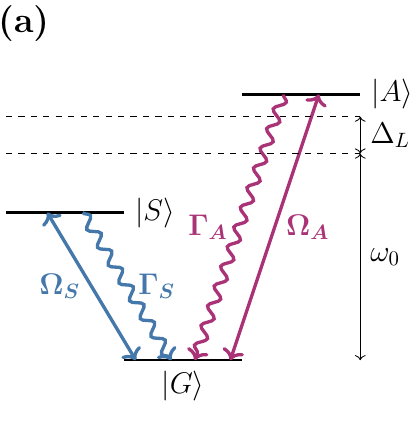}
\end{subfigure}
\hfill
\begin{subfigure}[b]{0.7\textwidth}
    \centering
    \includegraphics[scale=1.0]{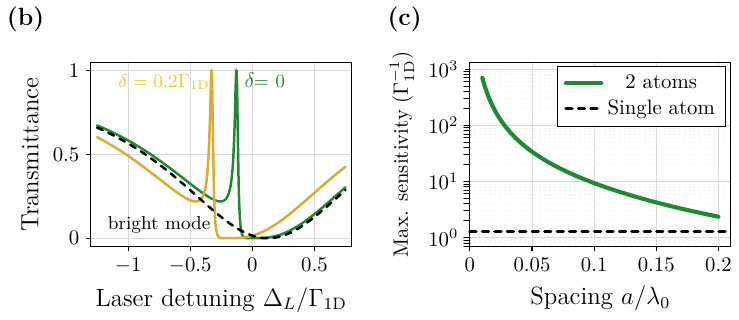}
\end{subfigure}

\caption{\textbf{
Two atoms coupled to a waveguide near the Dicke limit}. We consider the system driven by a weak field detuned $\Delta_L$ from the atom frequency $\omega_0$. (a) is a level diagram depicting the collective ground state $\ket{G}$, and the one-excitation eigenstates ---the antisymmetric $\ket{A}$ and symmetric $\ket{S}$ states--- with their decay rates and Rabi frequencies. (b) Transmittance as a function of the laser detuning. The green line is the transmittance of two atoms separated $0.04\lambda_0$, which is very similar to the transmittance only due to the bright mode (dashed line) except for a narrow feature caused by the subradiant mode. The yellow line represents the transmittance when an additional detuning $\delta=0.2\Gamma_{1D}$ is applied to both atoms. The narrow peak provides high sensitivity to global detunings. (c) Maximum of the sensitivity with respect to the laser detuning as a function of spacing $a$ between atoms. For comparison, the dashed line shows the maximum sensitivity attainable by a single atom.
}
\label{fig:III.1}
\end{figure*}

In this section, we present two distinct mechanisms that enable high sensitivity of the transmittance to variations in the laser and atomic transition frequencies. Due to collective effects, subradiant states with suppressed decay give rise to sharp features in the transmission spectrum. As a result, the transmittance close to this narrow feature becomes highly sensitive to small changes in the laser detuning $\Delta_L$ and to external global frequency shifts  $\delta_i \equiv \delta$ on the atoms, as illustrated in Fig.~\ref{fig:III.1}(b).

This enhanced sensitivity has several potential applications. First, it could be harnessed for atomic clock implementations, where shifts of the laser frequency are detected via changes in transmittance rather than by monitoring single-atom populations, as is typically done. Second, if the laser frequency is held fixed, a global frequency offset would simply shift the transmission profile, enabling precise readout of the detuning through transmission measurements.

\subsection{Enhanced sensitivity via subradiant states in waveguide QED}
 
As a first mechanism to measure global frequency shifts, we consider the case of two atoms coupled to a one-dimensional waveguide. The emitters are separated by a distance $a$, and driven by an incident electric field detuned $\Delta_L = \omega_L - \omega_0$ from the atomic transition frequency. For simplicity, we assume that the driving laser couples resonantly to the guided mode of the waveguide, i.e. $\omega_L = \omega_p$, and that no additional sources of detuning are present ($\delta_i = 0$). Then, the single-excitation eigenstates of the effective Hamiltonian in Eq.~(\ref{eq: effective_Ham}) are the symmetric and antisymmetric superposition states,
\[
\ket{S} = \frac{1}{\sqrt{2}}(\ket{e_1} + \ket{e_2}), \quad
\ket{A} = \frac{1}{\sqrt{2}}(\ket{e_1} - \ket{e_2}),
\]
where $\ket{e_i}$ denotes the excited state of atom $i$. Their eigenvalues (including collective decay rates and energy shifts) are 
\begin{subequations}
\begin{align}
    \lambda_S &=  \frac{\Gamma_{1D}}{2}\sin(k_p a) - i\Gamma_{1D} \cos^2\left(\frac{k_p a}{2}\right),\\
    \lambda_A &=  -\frac{\Gamma_{1D}}{2}\sin(k_p a) - i\Gamma_{1D} \sin^2\left(\frac{k_p a}{2}\right),
\end{align}
\end{subequations}
and depend sensitively on the interatomic spacing $a$. Owing to the periodicity of light-induced interactions in the waveguide, the system exhibits identical behavior for all separations $a + n\lambda_0$ with $a \in [0, \lambda_0]$ and $n \in \mathbb{N}$. For $a = n\lambda_0$, with $n \in \mathbb{N}$, the system reaches the Dicke limit: the symmetric state becomes superradiant with a decay rate $\Gamma_S = 2\Gamma_{1D}$, while the antisymmetric state is perfectly dark ($\Gamma_A=0$) and does not decay. In this case, both collective modes exhibit zero energy shifts.\footnote{For $a = (2n + 1)\lambda_0/2$ with $n \in \mathbb{Z}$, the system also reaches a Dicke-like limit, but with the radiative properties of the collective modes reversed: the symmetric state becomes dark, while the antisymmetric state becomes superradiant.} 

We work near the Dicke limit, where the antisymmetric state acquires a small but nonzero decay rate and is no longer perfectly dark, as illustrated in the level diagram of Fig.~\ref{fig:III.1}(a). In this regime, the symmetric and antisymmetric states are respectively driven with strengths $\Omega_S = \Omega_0 \cos(k_p a/2)$ and $\Omega_A = \Omega_0 \sin(k_p a/2)$, leading to the steady-state amplitudes [see Eq.~(\ref{eq:steady_state})]
\begin{subequations}\label{eq:pops_2_atoms}
\begin{align}
    \langle \hat{\sigma}_S \rangle &= \frac{\langle \hat{\sigma}_1 \rangle + \langle \hat{\sigma}_2 \rangle}{2} = \frac{\Omega_S}{\lambda_S - \Delta_L}, \\
    \langle \hat{\sigma}_A \rangle &= \frac{\langle \hat{\sigma}_1 \rangle - \langle \hat{\sigma}_2 \rangle}{2} = \frac{\Omega_A}{\lambda_A - \Delta_L}.
\end{align}
\end{subequations}

The corresponding transmission and reflection coefficients are given by
\begin{subequations}
\begin{align}
    t &= 1 + i\Gamma_{1D} \left[ \frac{\cos^2(k_p a/2)}{\lambda_S - \Delta_L} + \frac{\sin^2(k_p a/2)}{\lambda_A - \Delta_L} \right], \\
    r &= i\Gamma_{1D} \left[ \frac{\cos^2(k_p a/2)}{\lambda_S - \Delta_L} - \frac{\sin^2(k_p a/2)}{\lambda_A - \Delta_L} \right].
\end{align}
\end{subequations}

Note from Eq.~(\ref{eq:pops_2_atoms}) that the population in each collective state follows a Lorentzian profile centered at $J_\lambda$ with width $\Gamma_\lambda$. Far from resonance with the subradiant state $|A\rangle$, the superradiant state $|S\rangle$ is predominantly populated, resulting in a broad transmission spectrum that resembles that of a single emitter but with twice the linewidth [see Fig.~\ref{fig:III.1}(b)].

Importantly, the steady-state amplitudes and the resulting transmission spectrum are significantly modified near resonance with the subradiant state. In particular, at $\Delta_L = J_A$, the steady-state amplitude of the subradiant mode is given by $\langle \hat{\sigma}_A \rangle = -2i\Omega_A / \Gamma_A$. Although the Rabi frequency $\Omega_A$ decreases as the system approaches the Dicke limit, the decay rate $\Gamma_A$ decreases more rapidly. As a result, the steady-state amplitude $\langle \hat{\sigma}_A \rangle$ can become substantially larger than that of the superradiant state, which remains approximately constant, $\langle \hat{\sigma}_S \rangle \approx 2\Omega_S / \Gamma_S$. 

Due to the dominant population of the subradiant state, the fields scattered by both collective modes become comparable and interfere. One can demonstrate analytically that the reflected fields interfere perfectly destructively at
\begin{equation}
    \Delta_L = -\frac{\Gamma_{1D}}{2}\tan(k_p a),
\end{equation}
resulting in perfect transmission. As illustrated in Fig.~\ref{fig:III.1}(b), this leads to a narrow transparency window of width $\propto \Gamma_A$, where the transmittance rises sharply from near zero to unity over a small frequency range. This sharp feature yields high sensitivity to global frequency shifts, quantified by $|dT/d\Delta_L|$. We define the maximum sensitivity as the value optimized over $\Delta_L$ and show it in Fig.~\ref{fig:III.1}(c). As the interatomic spacing decreases and approaches the Dicke limit, the subradiant decay rate—and thus the width of the transparency window—decreases, leading to greater sensitivity. In this limit, the system also becomes increasingly vulnerable to imperfections, as we discuss in Sec.~\ref{sec:non-perfect_coupling} and Sec.~\ref{sec:atom_motion}, and the timescale to reach steady state increases.

\subsection{Enhanced sensitivity via selective coupling to dark states}\label{subsec:III.B} 

\begin{figure*}
\centering
\begin{subfigure}{0.29\textwidth}
\centering
    \includegraphics[scale=0.70]{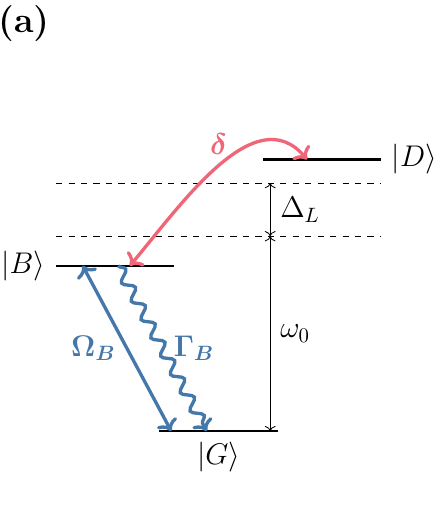}
\end{subfigure}
\hfill
\begin{subfigure}{0.7\textwidth}
    \centering
    \includegraphics[scale=1.0]{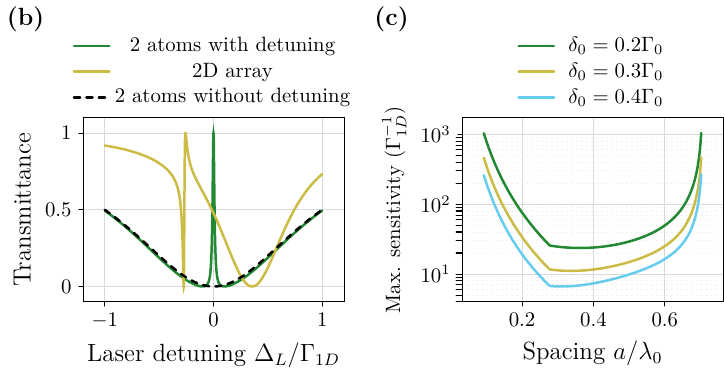}
\end{subfigure}
\caption{\textbf{Probing dark states via spatially dependent detunings}. We consider systems that have a perfectly dark state, such as two atoms coupled to a waveguide in the Dicke limit and an infinite 2D array in free space. In both cases, as depicted in the diagram (a), the dark mode $\ket{D}$ will be coupled to a bright mode $\ket{B}$ by applying a control space-dependent detuning $\delta_0$. The bright mode has a decay rate $\Gamma_B$ and Rabi frequency $\Omega_B$. $\ket{G}$ is the collective ground state. (b) Transmittance as a function of the laser detuning for two atoms coupled to a waveguide in the Dicke limit with detunings $\pm\delta_0$ (in green), and for a 2D array in free-space with spacing $a = 0.55\lambda_0$ and periodic detuning $\delta_0(\mbf{r}) = \delta_0 \cos(\pi/a \cdot (x+y))$ (in yellow). We choose $\delta_0 = 0.1\Gamma_0$. The dashed line is the transmittance of the two atoms in the Dicke limit without detuning. (c) Maximum sensitivity of an infinite 2D array, optimized over all laser detunings $\Delta_L$, as a function of the lattice spacing, shown for different values of the detuning amplitude.
}
\label{fig:III.2}
\end{figure*}

We now consider configurations that possess a perfectly dark state, which cannot be populated by the incident light alone. Two such examples are (i) an infinite two-dimensional (2D) atomic array in free space driven by a perpendicular laser field, and (ii) two atoms coupled to a waveguide in the Dicke limit, $a = n\lambda_0/2$, with $n \in \mathbb{N}$. In both cases, the incident light excites only a single collective mode, which we refer to as the bright mode $|B\rangle$. This mode corresponds to the symmetric state $\ket{S}$ in the two-atom setup, and to the collective spin wave with zero quasi-momentum, $\mathbf{k} = \mathbf{0}$, in the 2D array, as defined in Eq.~(\ref{eq:momentum_basis}). As a result, the transmittance spectrum exhibits the same profile as that of a single atom coupled to a waveguide, with a width determined by the decay rate of the bright mode. This spectrum lacks narrow features and therefore does not exhibit enhanced sensitivity.

To enhance sensitivity, we introduce a periodic control detuning that couples the bright mode to a dark mode of the system, as illustrated in Fig.~\ref{fig:III.2}(a). In the case of two atoms coupled to a waveguide, it is sufficient to apply an antisymmetric detuning $\delta_1 = - \delta_2 = \delta_0$ to populate the dark (antisymmetric) state $\ket{A}$. For the 2D array, we implement a position-dependent detuning of the form $\delta_0(\mathbf{r}) = \delta_0 \cos(\mathbf{k} \cdot \mathbf{r})$ with quasi-momentum $\mathbf{k} = (\pi/a, \pi/a)$, which couples the bright mode $\mathbf{k} = \mathbf{0}$ to the dark mode with quasi-momentum $\mathbf{k} = (\pi/a, \pi/a)$~\cite{rubies-bigorda_photon_2022}~\footnote{The idea of our protocol is to couple the bright mode to a single dark mode. In this work, we chose the dark mode with momentum $\mathbf{k} = (\pi/a, \pi/a)$, but it can equally be applied to those with momenta $\mathbf{k} = (\pi/a, 0)$ or $\mathbf{k} = (0, \pi/a)$.}
, defined in Eq.~(\ref{eq:momentum_basis}), that lies at the corner of the Brillouin zone of the square lattice and is perfectly dark if the inter-emitter spacing satisfies $a < \lambda_0 / \sqrt{2}$.

Under the position-dependent detuning, the steady-state amplitudes of the bright and dark modes are given by
\begin{subequations}\label{eq:pops_2D_array}
\begin{align}
    \langle \hat{\sigma}_B \rangle = \frac{\sqrt{N} \Omega_0 (\lambda_D - \Delta_L)}{(\lambda_B-\Delta_L)(\lambda_D-\Delta_L)-\delta_0^2}, \\
    \langle \hat{\sigma}_D \rangle= \frac{\sqrt{N} \Omega_0 \delta_0}{(\lambda_B-\Delta_L)(\lambda_D-\Delta_L)-\delta_0^2},
\end{align}
\end{subequations}
where $N$ is the number of atoms, and \( \lambda_B = J_B - i \Gamma_B/2 \) and \( \lambda_D = J_D \) are the eigenvalues of the effective Hamiltonian in Eq.~(\ref{eq: effective_Ham})---defined without the detuning modulation---corresponding to the bright and dark modes, respectively. In the case of two atoms coupled to a waveguide in the Dicke limit, both energy shifts vanish, $J_B = J_D = 0$. In contrast, for the 2D array, the energy shifts and bright decay rate vary with spacing $a$.

Since only the bright mode radiates, the transmission coefficient takes a simple form
\begin{equation}\label{eq:1}
    t = 1 + \frac{i}{2} \Gamma_0 \frac{\lambda_D - \Delta_L}{(\lambda_B - \Delta_L)(\lambda_D - \Delta_L) - \delta_0^2}.
\end{equation}
When the incident light is exactly resonant with the dark mode, i.e., $\Delta_L = J_D$, the drive and the coupling between modes interfere destructively, resulting in zero population of the bright mode. Then, only the subradiant state is excited, such that no light is absorbed and the transmittance reaches unity [see Fig.~\ref{fig:III.2}(b)]. Conversely, the transmission vanishes when the laser is resonant with one of the new eigenmodes of the system in the presence of the detuning pattern. These modes are obtained by diagonalizing the full effective Hamiltonian $\hat{H}_\mathrm{eff} + \hat{H}_\mathrm{detuning}$, and occur at the frequencies
\begin{equation}\label{eq:III.2.1}
    \Delta_L = \frac{1}{2} \left[J_B + J_D \pm \sqrt{(J_B - J_D)^2 + 4\delta_0^2} \right].
\end{equation}
In the regime of small detuning strength \( \delta_0 \), these transmission zeros are located near the bright ($\Delta_L \approx J_B$) and dark ($\Delta_L \approx J_D$) mode resonances. Notably, one of the points of total reflection lies in close proximity to the point of perfect transmittance, giving rise to the sharp feature presented in Fig.~\ref{fig:III.2}(b) and thereby enhancing sensitivity to frequency shifts.

For the case of two atoms coupled to a waveguide in the Dicke limit, where $J_B = J_D = 0$, the transmittance spectrum is symmetric [green trace in Fig.~\ref{fig:III.2}(b)] and the narrow peak is centered at $\Delta_L = 0$. As the detuning strength $\delta_0$ decreases, the transmission zeros move progressively closer to the peak, narrowing the transparency window and increasing sensitivity.

In contrast, for the 2D array, the bright and dark modes are no longer degenerate, resulting in an asymmetric transmission spectrum (yellow trace in Fig.~\ref{fig:III.2}(b)). The sharpest transition from zero to unit transmission occurs near $\Delta_L = J_D$, within a frequency window of approximately $\delta_0^2 / |J_B - J_D|$. Consequently, the sensitivity increases both as $\delta_0$ decreases and as the energy difference between the bright and dark modes grows, as illustrated in Fig.~\ref{fig:III.2}(c). This enhancement is particularly pronounced near the spacings $a \ll \lambda_0$ and $a = \lambda_0/\sqrt{2}$, where the frequency difference $|J_B - J_D|$ diverges.

\section{Site-resolved metrology}\label{sec:site-resolved_metrology}

In the previous section, we harnessed the dependence of the transmittance on the laser detuning $\Delta_L$
to detect changes in the laser frequency or global atomic detunings.
More generally, the transmittance profile depends on the positions of the atoms and their individual detunings, and thereby also allows for site-resolved metrology of these magnitudes.
In this section, we examine two complementary scenarios: (i) when the atomic positions are known, we aim to infer their detunings; and (ii) when the detunings are known, we seek to determine the atomic positions. We show that both tasks can be accomplished in an array coupled to a one-dimensional waveguide.

Physically, this is possible because, in an array of atoms coupled to a 1D waveguide, all collective modes can be individually addressed. Each mode responds primarily within a frequency window of width comparable to its linewidth and centered at its eigenfrequency. Importantly, each mode is sensitive to perturbations with a specific spatial periodicity. Therefore, by measuring the transmittance over a sufficiently broad frequency range to probe all modes, one can detect and reconstruct any arbitrary spatial perturbation in the atomic detunings or positions, thereby quantifying the disturbance experienced by each atom.

Mathematically, to be able to quantify any space-dependent perturbation, we will sample the transmittance spectrum at $M$ frequency points, with $M \geq N$ the number of atoms. In this case, the Jacobian $J_T$ is an $M \times N$ matrix with elements equal to the derivative of the transmittance with respect to the perturbation ---detunings or atomic positions--- at a certain frequency, $dT(\omega_i)/d \delta_j$ or $dT(\omega_i)/d r_j$. If the Jacobian matrix has full rank $= N$, the function is locally injective~\cite{SpivakMichael1965Coma, rudin_principles_1976}, implying that small perturbations can be inferred by sampling the transmittance spectrum.

However, full rank alone does not guarantee numerical stability: if the Jacobian is ill-conditioned, small measurement errors in transmittance can lead to large uncertainties in the inferred atomic parameters. We therefore analyze the condition number, $\kappa(J_T)$, defined as the ratio between the largest and smallest singular values of the Jacobian. A small condition number (close to $1$) indicates good numerical stability, while a large $\kappa(J_T)$ signals sensitivity to noise and difficulty to trace back the perturbation from the transmittance data. Note that, in our system, the condition number is generally large due to the disparity in sensitivity between superradiant and strongly subradiant modes.

In the first scenario, where the atomic positions are known and we aim to infer their detunings, the transmittance is generally not injective near zero detuning. This arises from the mirror symmetry of the array, which causes any detuning configuration $(\delta_1, \delta_2, \dots, \delta_N)$ and its mirror $(\delta_N, \delta_{N-1}, \dots, \delta_1)$ to produce an identical transmittance spectrum. To break this symmetry, the array must be initialized with a spatially dependent, asymmetric control detuning $\delta_0(r)$. Under this condition, a local one-to-one mapping from the transmittance to the detunings can be established, allowing small deviations from the known control detuning to be accurately quantified.

In the second scenario, we infer the atomic positions assuming the detunings are known. This approach is useful in experiments where the atoms deviate from an ideal array, providing an alternative to conventional fluorescence-based calibration methods. As in the previous case, a spatially varying, asymmetric detuning must be applied to break the array’s translational and mirror symmetries, ensuring a one-to-one mapping from the transmittance to the atomic positions.

To quantify the system’s sensitivity, we consider the norm of the gradient of the transmittance with respect to the atomic detunings or positions, $\lVert \nabla_{\mathbf{\delta}} T\rVert$ or $\lVert \nabla_{\mathbf{r}}T\rVert$. This norm depends on the probe frequency $\Delta_L$, since different frequency points probe different modes, each sensitive to perturbations with distinct spatial periodicities. To capture the system’s response to arbitrary spatial perturbations, a natural measure of sensitivity is the integral over all frequencies, $\int_{-\infty}^{+\infty} \lVert\nabla T\rVert d\Delta_L$.

\begin{figure}
    \centering
    \includegraphics[scale=1.0]{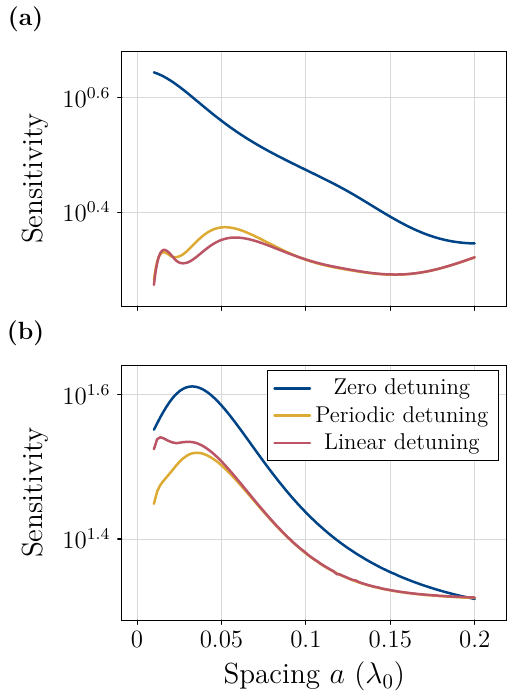}
    \caption{\textbf{Sensitivity to spatially dependent perturbations}.  
We demonstrate that applying an asymmetric control detuning enables site-resolved sensing in atomic arrays coupled to a waveguide. Panels (a) and (b) present the sensitivity of a 4-atom chain to (a) spatially varying detunings and (b) atomic displacements, as a function of the lattice spacing $a$, for different choices of control detuning: no detuning, \( \delta_0(r) = 0 \) (blue); periodic detuning, \( \delta_0(r) = 0.1 \sin(\pi r / N a) \) (yellow); and linear detuning, \( \delta_0(r) = 0.1 / [(N{-}1)a] r\) (red). The case with no detuning is included for comparison, though it is not experimentally feasible because the transmittance spectrum is not injective and thus cannot uniquely resolve local perturbations.}

    \label{fig:IV}
\end{figure}

Fig.~\ref{fig:IV} illustrates the sensitivity of a four-atom array to external detunings [Fig.~\ref{fig:IV}(a)] and positional displacements [Fig.~\ref{fig:IV}(b)]. The sensitivity is plotted for distinct control detunings to show how much it depends on the specific detuning pattern that we choose.
The plots also show the case without detuning, $\delta_0(r) = 0$ (blue lines), for which the transmittance is not injective and the reconstruction cannot be performed. 
All other cases have been chosen to ensure a locally well-conditioned one-to-one mapping (with condition number $\kappa(J_T)\leq 10^4$). Experimentally, one could choose the known detuning that gives the best sensitivity and good injectivity for the specific spacing $a$ of the array.

The discussion in this section applies only to an atomic array coupled to a waveguide. In a 2D free-space array, the geometry prevents us from significantly populating many subradiant states. Consequently, their emitted fields barely interfere with those of the superradiant modes, reducing the information about site-dependent detunings that can be inferred from the transmittance spectrum.

\section{Experimental sources of error and performance}\label{sec:errors}

Thus far, our analysis has assumed idealized conditions. We now examine how key experimental imperfections impact the sensitivity of the protocol. In particular, we consider three main sources of error: imperfect coupling between the atoms and the guided mode of the waveguide, atomic motion, and missing atoms in the array. Even in the presence of these imperfections, the protocol preserves enhanced sensitivity, demonstrating its robustness under realistic experimental conditions. We also evaluate the achievable frequency precision of the scheme and place it in context with leading atomic-clock performance.

\subsection{Effect of imperfect coupling between atoms and waveguide}\label{sec:non-perfect_coupling}

Up to this point, we have considered perfect coupling of the atomic array to the guided mode. We now account for uncorrelated decay into free space and unguided waveguide modes at rate $\Gamma'$ by adding the Lindbladian term 
$(\Gamma'/2) \sum_j \left( 2 \hat{\sigma}_j \hat{\rho} \hat{\sigma}_j^\dagger - \hat{\sigma}_j^\dagger \hat{\sigma}_j \hat{\rho} - \hat{\rho} \hat{\sigma}_j^\dagger \hat{\sigma}_j \right)$. Adding uncorrelated decay leaves the collective eigenstates unchanged and increases their decay rates by $\Gamma'$, leading to broader transmission peaks. Additionally, uncorrelated decay reduces the transmittance and reflection peaks, as a fraction of the absorbed light is irreversibly scattered into non-guided modes (such that $T+R < 1$).

\begin{figure}
    \centering
    \includegraphics[scale=1.01]{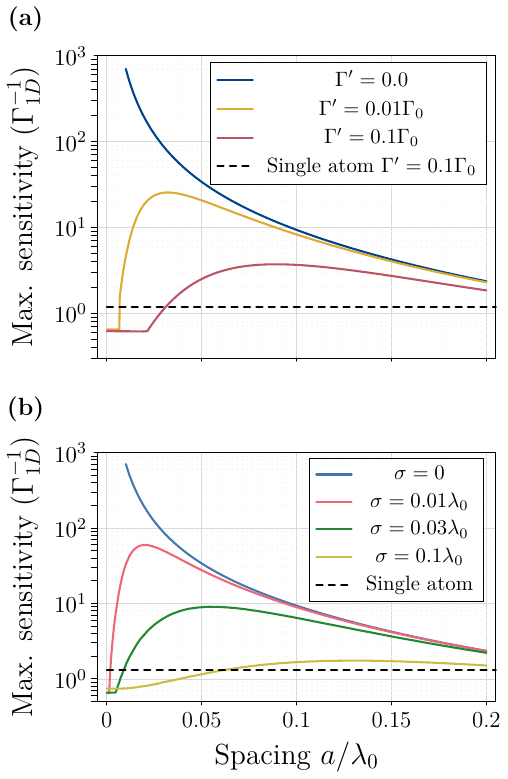}
    \caption{\textbf{Sensitivity of two atoms coupled to a waveguide including experimental errors}. Maximum sensitivity as a function of interatomic distance $a$. (a) Effect of imperfect atom-waveguide coupling for different values of the decay into non-guided modes, $\Gamma'$. (b) Effect of atomic motion for different values of zero-point motion, $\sigma$. The dashed lines correspond to the case of a single atom coupled to the waveguide.}
    \label{fig:non-perfect_coupling&atom_motion}
\end{figure}

Figure~\ref{fig:non-perfect_coupling&atom_motion}(a) shows that, in the case of two atoms, the sensitivity decreases with increasing $\Gamma'$. In contrast to the ideal case (of perfect coupling), sensitivity decreases as the interatomic separation $a \to 0$ for any fixed $\Gamma' > 0$. This behavior follows from the steady-state subradiant population given in Eq.~(\ref{eq:pops_2_atoms}). As $a \to 0$, the Rabi frequency of the subradiant mode vanishes, while the denominator $\lambda_- - \Delta_A$ is now bounded from below by $\Gamma' / 2$. As a result, the amplitude of the subradiant state tends to zero ($c_- \to 0$), such that the contribution of the subradiant state to the transmittance becomes negligible near the Dicke limit. The sensitivity is then dominated by the broader superradiant state, which is less sensitive than a single atom. Finally, it should be noted that, for any given $\Gamma'$, there exists an optimal interatomic spacing that maximizes sensitivity.

\subsection{Effect of atomic motion}\label{sec:atom_motion}

Another important source of experimental error is the motion of atoms within their trapping potential. Even when tightly confined by an optical lattice, atoms exhibit position fluctuations due to finite temperature.
Assuming a quadratic trap, the spatial wavefunction of cold atoms is well approximated by the ground state of a harmonic oscillator, characterized by its zero-point motion $\sigma$.

In Fig.~\ref{fig:non-perfect_coupling&atom_motion}(b), we consider two atoms coupled to a waveguide in the high-velocity limit, where each atom moves many times around its trap during the measurement. This limit can be modeled by averaging all magnitudes dependent on the atomic positions. Specifically, to compute the steady-state populations from Eq.~(\ref{eq:steady_state}), we average $J_{ij}$ and $\Gamma_{ij}$, which depend on positions via the Green's tensor, as well as the Rabi frequencies $\Omega_i$, whose phases are also position-dependent. Finally, in Eq.~(\ref{eq:electric_field}) for the electric field, the Green's tensor is also averaged.

Similarly to imperfect atom-waveguide coupling, atomic motion primarily broadens the subradiant collective modes of the array~\cite{Guimond2019,Rusconi2021,Rubies2025}. As a result, the sensitivity remains largely unchanged for large interatomic spacings, but decreases as the system approaches the Dicke limit. In this limit, the strongly subradiant mode is significantly broadened, such that the enhancement in sensitivity is lost.

\subsection{Effect of missing atoms}\label{Sec.V.C}

Another imperfection arises from the probabilistic loading of atoms into the optical lattice (or tweezers), which leaves a fraction of sites unoccupied.
Missing atoms break the spatial symmetry of the array and distort its collective optical response. As a result, the transmitted field becomes spatially inhomogeneous and strongly dependent on the detection point, so that the transmittance measured at a single point may not capture the full behavior. A practical solution is to average the transmitted signal over a surface or across multiple detection points, thereby recovering meaningful information about the response of the system.

\begin{figure}
    \centering
    \includegraphics[scale=1.0]{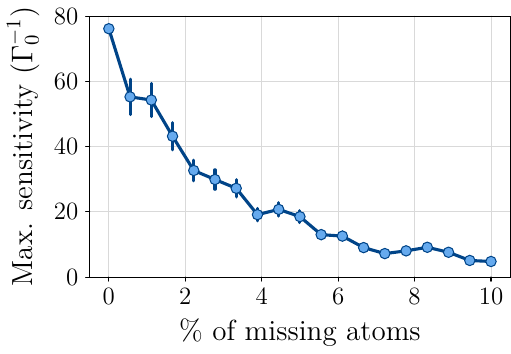}
    \caption{\textbf{Effect of missing atoms on sensitivity}. Average maximum sensitivity of a $10 \times 10$ array in free-space with spacing $a = 0.5\lambda_0$ and control detuning $0.1\lambda_0 \cos(\mbf{k} \cdot \mbf{r})$. The sensitivity is averaged over all possible arrays with a given percentage of atoms missing. The incident field is a Gaussian beam with beam waist $w_0 = 0.3\sqrt{N}a$ and the transmittance is sampled at $31$ points uniformly distributed within a circle of radius $1.2w_0$, centered along the axis perpendicular to the array at a distance $1.1\lambda_0$.}
    \label{fig:missing_atoms}
\end{figure}

In Fig.~\ref{fig:missing_atoms}, we study the sensitivity of a $10 \times 10$ array in free-space with a spacing $a = 0.5\lambda_0$ and a periodic detuning $\delta_0(\mbf{r}) = 0.1\Gamma_0 \cos(\mbf{k} \cdot \mbf{r})$, as a function of the percentage of missing atoms. Since the transmittance profile depends on the specific locations of the missing atoms, each data point in the plot represents the average over many realizations with randomly removed atoms. An increasing number of missing atoms reduces the symmetry of the emerging collective effects and, hence, sensitivity.

\subsection{Precision estimate and comparison}

After addressing the experimental feasibility of our proposal, we now estimate the achievable precision in detecting frequency shifts and compare it with the leading frequency-standard platforms.

The frequency uncertainty can be approximated as
\begin{equation}
    \Delta \omega \approx \frac{\Delta P}{|dP/d\omega|_{\mathrm{max}}},
\end{equation}
where $P$ is the incident power and $\Delta P$ its uncertainty. The slope is set by the subradiant feature, $|dP/d\omega|_{\mathrm{max}} \sim P / \Gamma_\mathrm{sub}$, with $\Gamma_\mathrm{sub}$ the subradiant linewidth. A convenient estimate for the signal-to-noise ratio is $P / \Delta P \equiv \mathrm{SNR} \approx C\sqrt{\eta_{\mathrm{eff}} P \tau / (\hbar \omega_0)}$, where $P \tau/\hbar \omega_0$ is the number of photons during a measurement time of $\tau$. The detection efficiency $\eta_{\mathrm{eff}}$ and the experimental contrast $C$---defined as the fraction of the full transmitted-power range that can be effectively resolved despite technical limitations---are assumed to be of order unity.

The protocol operates in the low-excitation regime, $p = \langle \sigma_i^\dagger \sigma_i\rangle_{\mathrm{max}} = |\Omega / \Gamma_0|^2 \ll 1$, with $\Omega$ the Rabi frequency determined by the input power. Notably, this condition inherently limits the maximum incident power to $P \sim p N \hbar \omega_0 \Gamma_0$. Together with the expression for the signal-to-noise ratio, this yields the frequency uncertainty
\begin{equation}\label{eq:uncertainty_freq2}
    \Delta \omega \sim \frac{\Gamma_\mathrm{sub}}{\sqrt{p N \Gamma_0 \tau}}.
\end{equation}

To estimate the precision achievable with current systems, we examine the $D_2$ line of $^{87} \mathrm{Rb}$, previously used in subwavelength atomic array experiments~\cite{rui_subradiant_2020}. Taking $N = 1000$ atoms, an excited population per atom of $p = 0.01$, and a subradiant linewidth a hundred times lower than that of a single atom, $\Gamma_\mathrm{sub} = \Gamma_0 / 100$, the Allan deviation is $\sim 10^{-14}$ for $\tau = 1$s. Here $\tau$ refers to the measurement (photon-collection) time once the system has reached steady state. Note that it differs from the typical integration time in atomic clocks, which refers to the total averaging time over which the clock frequency is tracked. For comparison, leading optical clocks reach $10^{-16} - 10^{-17}$ at one second and $10^{-18} - 10^{-19}$ over hours~\cite{bothwell_jila_2019, oelker_demonstration_2019}. This exceptional precision is partly enabled by the long integration time, which we can in principle also realize: after preparing the array and allowing the system to reach a steady state, we measure the transmittance and adjust the laser frequency accordingly. Notably, this procedure naturally supports continuous monitoring. Once atom loss or heating reaches a level at which the array’s sensitivity is noticeably reduced, the array can be re-prepared to recover optimal operating conditions. A second key ingredient underlying the high precision of atomic clocks is the use of ultranarrow transitions, such as the millihertz line in strontium. If operated on such narrow transitions, Eq.~(\ref{eq:uncertainty_freq2}) indicates that subwavelength arrays may reach the $10^{-19}$ level with an measurement time of $\tau = 1\,\mathrm{s}$. Such small decay rates would increase the time to reach steady state and might also constrain the maximum allowable excitation.

\section{Conclusion and Outlook}

In summary, we have demonstrated that the narrow linewidths of subradiant states give rise to sharp features in the transmittance spectrum, which can be harnessed for precision sensing.
Specifically, we demonstrated that variations in laser frequency or global detuning can be detected using atomic arrays coupled to a waveguide (either near the Dicke limit or exactly in the Dicke limit with a control detuning) as well as with two-dimensional free-space arrays subject to a control detuning.
We further analyzed the possibility of site-resolved metrology of detunings and atomic positions. Crucially, we have also demonstrated that these systems provide enhanced sensitivity even in the presence of realistic experimental imperfections. In addition to demonstrating robustness, we have estimated the frequency precision achievable with this protocol and benchmark it against leading optical clocks. Existing realizations of subwavelength arrays already enable meaningful precision.
Their performance could be further improved, potentially matching that of leading atomic clocks, through the use of ultranarrow atomic transitions.

The results presented in this work are directly accessible in current experimental platforms. Subwavelength atomic arrays have been realized in free space using optical lattices~\cite{rui_subradiant_2020, rubies_thesis_mit} and optical tweezers~\cite{holman2024trappingsingleatomsmetasurface}. Alternatively, arrays of emitters coupled to a waveguide can be engineered using superconducting qubits coupled to a common transmission line~\cite{Painter_QO,Kirchmair,WillOliver_unidirectionalemission} or with atoms and quantum dots placed near nanophotonic waveguides~\cite{Goban_Waveguide,Goban_lightmatter,Collective_QD}.

Future research could explore protocols that go beyond the low-excitation limit, allowing stronger probe fields for faster and more robust imaging.
Moreover, our analysis uses only the transmitted intensity as the sensing observable; incorporating phase-resolved detection could provide additional signal channels and improve sensitivity. 
Finally, the time required to reach steady state depends on the initial state of the array. Identifying states that minimize this time can therefore improve practical measurement rates.

\vspace{1em}
\noindent \emph{Acknowledgments -} D.Z.-B. acknowledges support from the CFIS Mobility Program ---through Fundació Privada Mir-Puig, CFIS partners, and donors of the crowdfunding initiative--- and the Erasmus+ program, as well as additional support from AGAUR (MOBINT-MIF) and Banco Santander. O.R.-B. acknowledges support from Fundación Mauricio y Carlota Botton and from Fundació Bancaria “la Caixa” (LCF/BQ/AA18/11680093). S.F.Y. acknowledge support from NSF via PHY-
2207972, the CUA PFC PHY-2317134, and QuSeC-TAQS OMA-2326787 in addition to AFOSR FA9550-24-1-0311.

\bibliographystyle{apsrev4-2}
\bibliography{refs}


\clearpage

\appendix

\section{Non-smooth behavior of the maximum sensitivity of a 2D array as a function of the spacing}

\begin{figure}[h!]
    \includegraphics[scale=0.99]{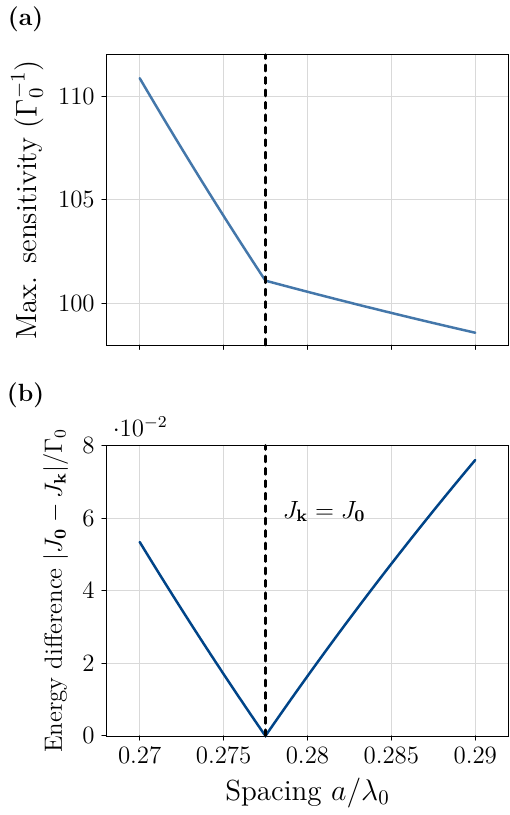}
    \caption{\textbf{Discontinuity in the sensitivity of an infinite 2D array as a function of the spacing}. (a) Maximum sensitivity of an infinite 2D atomic array with a detuning $\delta_0(\mathbf{r}) = 0.1 \Gamma_0 \cos(\mathbf{k} \cdot \mathbf{r})$, plotted as a function of the lattice spacing $a$. A non-smooth feature appears near $a \approx 0.28\lambda_0$. (b) Absolute value of the frequency difference between the bright and dark collective modes as a function of the spacing. The slope of this energy difference changes abruptly at the crossing point, where both modes become degenerate. Since the sensitivity depends strongly on the energy separation between these modes, it inherits the non-analytic behavior at this point.
}
    \label{fig:why_spike}
\end{figure}

In Sec.~\ref{subsec:III.B}, we saw that one of the ways to measure precisely the laser frequency or global frequency shifts was with a 2D array in free space with a periodic control detuning. Specifically, Fig.~\ref{fig:III.2}(c) depicts the maximum sensitivity of an infinite 2D array as a function of distance between emitters. One can see near the point $a = 0.28\lambda_0$ there is a discontinuity in the derivative of the sensitivity, which is shown in more detail in Fig.~\ref{fig:why_spike}(a). Here, we explain this non-smooth behavior.

Firstly, we can see that the maximum sensitivity (taken over all frequencies $\Delta_L$) depends on the positive energy difference between the bright and dark modes, $|J_0-J_k|$, with a simple argument. Note that varying the frequencies of the modes by the same constant is just a translation of the transmittance spectrum and will not affect the maximum sensitivity, thus, it only depends on the energy difference between the modes, $J_0-J_k$. Moreover, if we change the sign of the frequencies of the modes, the transmittance spectrum is mirrored and the maximum sensitivity remains the same. Therefore, the maximum sensitivity only depends on the absolute value $|J_0-J_k|$. In fact, in Sec.~\ref{subsec:III.B}, we already mentioned that the sharpest transmittance transition from zero to unit occurs in a frequency range of width approximately equal to $\delta_0^2 / |J_0-J_k|$. Therefore, the maximum sensitivity is directly proportional to the energy difference $|J_0-J_k|$.

This dependence is relevant because it turns out that the derivative of the absolute value $|J_0-J_k|$ as a function of the spacing is discontinuous, as Fig.~\ref{fig:why_spike}(b) illustrates. That discontinuity is translated to the plot of the maximum sensitivity of a 2D array as a function of $a$.

\section{Sensitivity of a 2D array as a function of atom number}

In Sec.~\ref{subsec:III.B}, we studied an infinite 2D array, and although we did consider a finite array with missing atoms in Sec.~\ref{Sec.V.C}, the dependence of the sensitivity of a finite 2D array with the atom number is yet to be seen.

\begin{figure*}
    \includegraphics[scale=0.99]{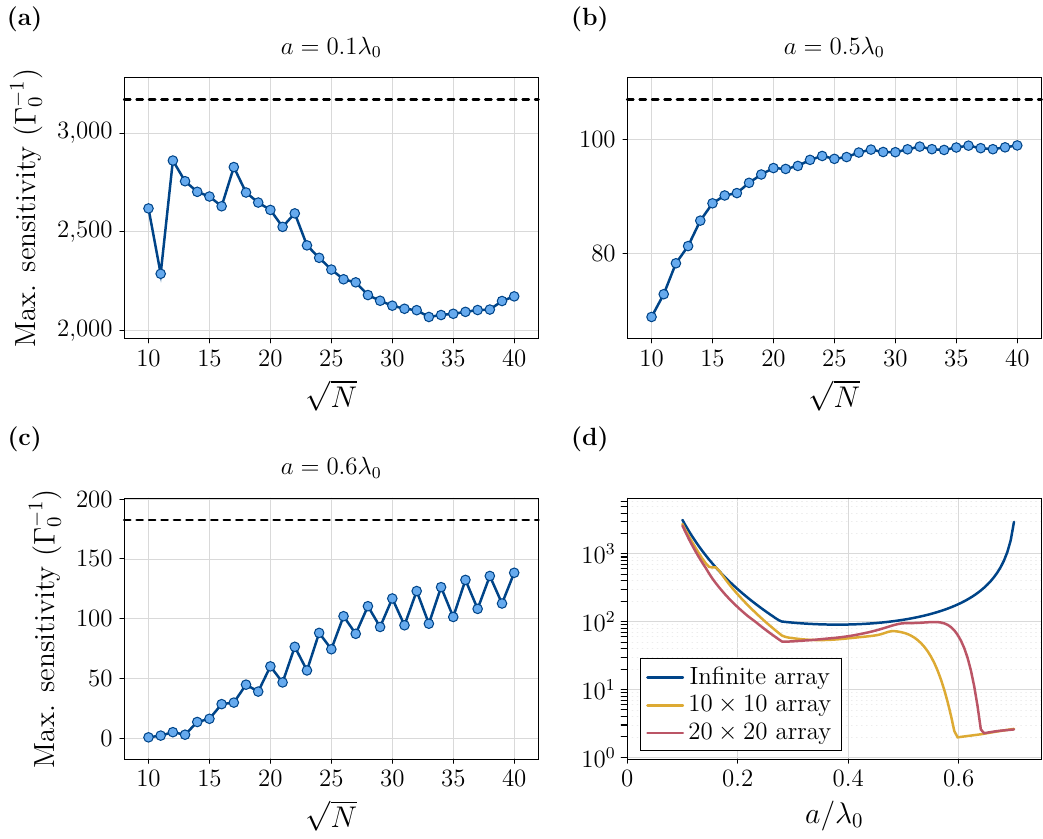}
    \captionof{figure}{\textbf{Sensitivity of a 2D array as a function of the atom number}. (a), (b) and (c) show the maximum sensitivity of a finite 2D array as a function of $\sqrt{N}$ ---the number of atoms per side of the lattice--- for three distinct spacings. The dashed line is the maximum sensitivity of the infinite 2D array with that spacing. (d) illustrates the maximum sensitivity as a function of the spacing for the infinite array, $10 \times 10$ and $20 \times 20$.}
    \label{fig:sensitivity_vs_N}
\end{figure*}

Fig.~\ref{fig:sensitivity_vs_N}(d) shows the maximum sensitivity of an infinite, a $10 \times 10$, and a $20 \times 20$ array for a detuning $\delta_0(\mbf{r}) = 0.1\Gamma_0 \cos(\mbf{k} \cdot \mbf{r})$. The sensitivity is always worse for a finite array, but close to the infinite case, except when approaching the light cone. Comparing the results between the $10 \times 10$ and the $20 \times 20$ array, they depend much on the specific spacing, and more atoms does not always ensure better sensitivity.

For a fixed distance, we would expect the sensitivity to grow and approach the sensitivity of the infinite array for the limit of many atoms, as we can see in Fig.~\ref{fig:sensitivity_vs_N}(b). However, in general, the sensitivity does not increase monotonously, as in Fig.~\ref{fig:sensitivity_vs_N}(c), where it has larger values for even number of atoms. In fact, in Fig.~\ref{fig:sensitivity_vs_N}(a), we cannot even see an upward trend yet, and there are some big oscillations for the smaller numbers of atoms.

\end{document}